\documentclass{elsart}
\usepackage{natbib}
\begin{document}

\runauthor{Cicero, Caesar and Vergil}

\begin{frontmatter}

\title{A Proposed Absolute Entropy of Near Extremal Kerr-Newman Black Hole}
\author{Hai Lin}
\address{Department of Physics, Peking University, P.R.China, 100871}
\thanks[Someone]{Email:hailin@mail.phy.pku.edu.cn} 

\begin{abstract}
    Some problems have been found as to the definition of entropy of black hole 
\(S = \frac{A}{4}\) being applied to the extremal Kerr-Newman case, which has
conflicts with the third law of thermodynamics. We have proposed a new
modification for the near extremal one, which not only obeys the third law, but also
does not conflict with other conclusions in black hole thermodynamics. Then
 we proved that the inner horizon has temperature \({T}_{-}^{}= \frac{{\kappa}_
{-}^{}}{2\pi }\) and proposed that the inner horizon contributes to the 
entropy of the near extremal one so that the entropy of it is assumed to be \(S = ({A}_{+}
^{}+{A}_{-}^{})/4\) and vanishes at absolute zero temperature.
\end{abstract}

\end{frontmatter}

\section{Introduction}   
    The thermodynamic properties of black hole have long been researched by
scientists. The radiation temperature was regarded as the surface
gravity at the outer horizon [1]. For a Schwarzschild black hole, it is       
\begin{equation}\label{fo:T}  
 T = \frac{\kappa }{2\pi }=\frac{1}{8\pi M}, 
\end{equation}                
where M is the mass of the black hole. And the entropy of black hole was
regarded as the area of the outer event horizon [2] :                         
\begin{equation}\label{fo:S}
  \  S= \frac{A}{4},
\end{equation}
\begin{equation}\label{fo:A}
\ A =8\pi \lfloor {M}_{}^{2}+M(  \sqrt{{M}_{}^{2}-{J}_{}^{2}/
{M}_{}^{2}-{Q}_{}^{2} }) -{Q}_{}^{2}/ 2\rfloor ,
\end{equation}
where  J, Q are the angular momentum and electric charges of black hole. 
    Also, scientists have established the four laws of black hole thermodynamics
relative to the four laws in thermodynamics [1,3,4]. And S. Hawking discovered the 
radiation of black hole; thus the temperature of black hole got a real significance.
The temperature of black hole has been widely researched these years but the 
study of entropy might not be sufficient [5-9]. 
    However,the entropy in formula (2) does not obey the third law of 
thermodynamics [10,11],which is Nerst's theorem: according to Planck gauge, the 
entropy of a system will vanishes when its temperature approaches to absolute zero.
Actually, according to formula (2) and (3), when \(T ={(\frac{\partial M}{\partial S})}_{J,Q}^{}\rightarrow0\)
 (the so called "extremal black hole" )
, we have \(S\rightarrow \pi  \sqrt{{Q}_{}^{4}+4{J}_{}^{2} }\ne 0\). So this
definition (\(S= \frac{A}{4}\) ) was unable to be applied to the extremal case , therefore is not
the Planck absolute entropy near zero temperature.
    In order to solve this problem, some scientists, for instance, Hawking, 
Teitelbom, and Gibbons etc.suggested to redefine the entropy of a near extremal black hole
[12-17], and thought it should be zero. Meanwhile, some scientists, such as
 Loranz, Hiscock, Zaslavskii etc. [18-23] insist on the former definition.Now it is 
still being discussed.
    However, we agree with the first group of scientists and redefine the absolute 
entropy of Kerr-Newman black hole near zero temperature, making it not only
obey the third law, but also be harmonious with other conclusions in black
hole thermodynamics.

\section{The Entropy of Near Extremal Kerr-Newmann Black Hole Derived From Thermodynamics 
}

    The entropy we want to find should at least accord with the following three
conditions:  1.	Obey the Bekenstein-Smarr function: 
\begin{equation}           
dM=TdS+\Omega dJ+VdQ ;
\end{equation}     
2.	Obey the limit relations between T and S  :  
            \begin{equation}           
 S\rightarrow 0 ( when  T\rightarrow 0);        
\end{equation}
  3.  The formula will be reduced to \( S = \frac{A}{4} \)    when\(J,Q\rightarrow 0\) . 
    Accordingly, by function (4)(condition 1), we have  
\begin{equation}   
T = { \left( \frac{\partial M }{\partial S } \right)s}_{J,Q}^{} = {
\left(\frac{\partial M }{\partial A }\right) }_{J,Q}^{}{\left( \frac{\partial A
}{\partial S }\right) }_{J,Q}^{}. 
\end{equation}      
    And from formula (3), we have
\begin{equation}   
  {\left( \frac{\partial M
}{\partial A }\right) }_{J,Q}^{} = \frac{1}{4} \left[ \frac{1}{8\pi M
}-{\left(  \frac{{Q}_{}^{4}}{4}+{J}_{}^{2}\right)}  \frac{8\pi }{M{A}_{}^{2}}
\right] .
\end{equation}  
 
    In order to accord with condition 3, we suppose the entropy has the form: 
\begin{equation}
S = \frac{A-{A}_{0}^{}}{4},
\end{equation}             
where \( {A}_{0}^{} \)    is a correction item, it may be the function of M ,J
,Q and vanishes when J and Q vanishes(condition 2).    
    From formula (8), we have 
\begin{equation}
{ \left(\frac{\partial A }{\partial S } \right)}_{J,Q}^{} = 4+ {\left(
\frac{\partial {A}_{0}^{}}  {\partial M }\right) }_{J,Q}^{}{
\left(\frac{\partial M }{\partial S } \right)}_{J,Q}^{}. 
\end{equation}      
    From equation (6) and (9), we have 
\begin{equation}
{\left( \frac{\partial A }{\partial S }\right) }_{J,Q}^{} = 4+T{\left(
\frac{\partial {A}_{0}^{} }{\partial M }\right) }_{J,Q}^{}. 
\end{equation}             
    Putting equation (7) and (10) into (6), we
have 
\begin{equation}
\frac{\partial {A}_{0}^{} }{\partial M } = \frac{32\pi
M{A}_{}^{2}}{{A}_{}^{2}-16{\pi }_{}^{2}({Q}_{}^{4}+4{J}_{}^{2})}- \frac{4}{T}. 
\end{equation}                           
     From formula (8) and in order to accord with the condition 2 , we get
when \( T\rightarrow 0, A\rightarrow {A}_{0}^{} \) . So \( {A}_{0}^{} \) equals to the area of
outer horizon at zero temperature. From formula (3), we have                 
\begin{equation}
{A}_{0}^{}=4\pi  \sqrt{{Q}_{}^{4}+4{J}_{}^{2} }.
\end{equation}                 
   So that                                  
\begin{equation}
\frac{\partial {A}_{0}^{}}{\partial M}=0.
\end{equation}
    From formulas (3), (8) and (12), we have                
\begin{equation}
S = 4\pi M \sqrt{{M}_{}^{2}-{J}_{}^{2}/ {M}_{}^{2}-{Q}_{}^{2} }    
(T\rightarrow 0). 
\end{equation}        
    From equation (11) and (13), we get 
\begin{equation}
T = \frac{{A}_{}^{2}-16{\pi }_{}^{2}({Q}_{}^{4}+4{J}_{}^{2})}{8\pi
M{A}_{}^{2}}= \frac{{A}_{}^{2}-{A}_{0}^{2}}{8\pi M{A}_{}^{2}},
\end{equation} 
which one can prove is exactly the Hawking temperature. When \( T\rightarrow 0,
\)    we have \( A\rightarrow {A}_{0}^{} \), so that\(S\rightarrow 0,\) it
obeys condition 2.And for an uncharged non-rotating black hole, formula (14)
will be reduced to formula (2).Thus we think formula (14) may be the absolute
Planck entropy of near extremal Kerr-Newman black hole approaching to zero
temperature.

\section{The Temperature, Heat Capacity and Radiation Power of Near Extremal Black Hole  }
    Formula (15) can be written in another form: 
\begin{equation}
T = \frac{1}{8\pi M}(1- \frac{{A}_{0}^{2}}{{A}_{}^{2}})       ,
\end{equation}                                                
witch will be reduced to formula (1) for an uncharged non-rotating black
hole.      
   From formula (16) we get the heat capacity 
\begin{equation}
{C}_{V}^{}={C}_{A}^{}={ \left(\frac{\partial M}{\partial T}\right) }_{A}^{}=-
\frac{8\pi {M}_{}^{2}}{\left(1- \frac{{A}_{0}^{2}}{{A}_{}^{2}}\right)},
\end{equation}     
which will also be reduced to  
\begin{equation}
{C}_{V}^{}=-8\pi {M}_{}^{2}
\end{equation}
for an uncharged non-rotating black hole . 
   Moreover, from formula (16) we get the radiation power 
\begin{equation}
P = \frac{\partial M}{\partial t} = \sigma A{T}_{}^{4} = \frac{\sigma
A}{{\left(8\pi\right)}_{}^{4}{M}_{}^{4}}{\left(1-
\frac{{A}_{0}^{2}}{{A}_{}^{2}} \right) }_{}^{4},  
\end{equation}             
where \( \sigma \) is the Stefan-Boltzman constant. For
an uncharged non-rotating black hole, we get
\begin{equation}
P = \frac{\sigma }{{\left(8\pi\right)}_{}^{3}{M}_{}^{2}}.
\end{equation}                                
  We can see from this section that the redefinition of the
entropy for the bear extremal Kerr-Newman black hole in section 2 does not conflict
to other thermodynamic properties of non-extremal, uncharged or non-rotating
ones .

\section{The Physical Interpretation of the Entropy of Near Extremal 
Kerr-Newman Black Hole}
   Now we would like to study the microcosmic mechanism of the  entropy
production of  near extremal Kerr-Newman black hole . If we choose the coordinates \(
{x}_{}^{\nu} = ( t, r, \theta ,\phi)\) , we have the Kerr-Newman metric
describing the geometry of a rotating charged black hole as follows:
\begin{eqnarray}
d{s}_{}^{2} = -\left(1- \frac{2Mr-{Q}_{}^{2}}{{\Sigma}_{}^{2}}\right)d{t}_{}^{2}
+ \frac{{\Sigma }_{}^{2}}{\Delta}d{r}_{}^{2}+{\Sigma }_{}^{2}d{\theta
}_{}^{2}  \nonumber \\      
+\left[({r}_{}^{2}+{a}_{}^{2}){\sin }_{}^{2}\theta +
\frac{(2Mr-{Q}_{}^{2}){a}_{}^{2}{\sin }_{}^{4}\theta }{{\Sigma }_{}^{2}}
\right]d{\phi }_{}^{2} \nonumber \\
-\frac{2(2Mr-{Q}_{}^{2})a{\sin }_{}^{2}\theta }{{\Sigma
}_{}^{2}}dtd\phi ,
\end{eqnarray}             
   
 in which we use the standard notations  
\begin{equation}
{\Sigma }_{}^{2} = {r}_{}^{2}+{a}_{}^{2}{\cos }_{}^{2}\theta , \Delta
= {r}_{}^{2}-2Mr+{a}_{}^{2}+{Q}_{}^{2},  a=J/M.  \end{equation}         
    The surface gravity at the outer (\(r={r}_{+}^{}=M+ \sqrt{{M}_{}^{2}-{a}_{}^{2}-{Q}_{}^{2} } )\) 
or inner (\(r={r}_{-}^{} = M-\sqrt{{M}_{}^{2}-{a}_{}^{2}-{Q}_{}^{2} } )\)  horizon is the limit of the
intrinsic acceleration b  multiplied by the redshift factor\(-
\sqrt{{g}_{00}^{} } \)  :           
\begin{equation}
{\kappa }_{\pm}^{} = \lim_{r\rightarrow {r}_{\pm}^{}}(-b \sqrt{{g}_{00}^{} }
) = \frac{{r}_{+}^{}-{r}_{-}^{}}{2({r}_{\pm}^{2}+{a}_{}^{2})} . 
\end{equation}      
    The charged particle in curved space-time obeys Klein-Gordon equation:
\begin{equation}
 \frac{1}{ \sqrt{-g } }\left( \frac{\partial }{\partial {x}_{
}^{\mu}}-ie{A}_{\mu}^{}\right ) \left[ \sqrt{-g }{g}_{}^{\mu \nu } \left(
\frac{\partial }{\partial {x}_{}^{\nu}}-ie{A}_{\nu}^{}\right )\Phi \right] =
{\mu }_{0}^{2}\Phi ,
\end{equation}    
where\({\mu }_{0}^{}\) and e are the static masss and electric charge of the
particle, and \({A}_{\mu}^{}\)  is the vector potential of the electromagnetic
field [9]:  
\begin{equation}
{A}_{\mu}^{}=- \frac{Qr}{\Sigma }(1,0,0,-a{\sin }_{}^{2}\theta  ) .
\end{equation}      
    Using the metrics (21) and formula (25), the Klein-Gordon equation (24)
is reduced to 
\begin{eqnarray}
\left[\frac{{({r}_{}^{2}+{a}_{}^{2})}_{}^{2}}{\Delta }-{a}_{}^{2}{\sin
}_{}^{2}\theta\right]\frac{{\partial }_{}^{2}}{\partial
{t}_{}^{2}}\Phi - 2a\left[1-\frac{({r}_{}^{2}+{a}_{}^{2})}{\Delta }\right]
\frac{\partial }{\partial t} \frac{\partial }{\partial \phi }\Phi
\nonumber \\
-\frac{\partial}{\partial r}\Delta \frac{\partial }{\partial r}\Phi 
-\frac{1}{\sin \theta} \frac{\partial }{\partial \theta } \sin
\theta  \frac{\partial }{\partial \theta }\Phi + \left(\frac{{a}_{}^{2}}{\Delta
}-\frac{1}{{\sin }_{}^{2}\theta }\right)\frac{{\partial }_{}^{2}}{\partial
{\phi}_{}^{2}}\Phi       \nonumber \\ 
 +2ie \frac{Qr}{\Delta }\left[a \frac{\partial}{\partial \phi }+
({r}_{}^{2}+{a}_{}^{2}) \frac{\partial }{\partial t}\right]\Phi +\left({\mu
}_{0}^{2}{\Sigma }_{}^{2}- \frac{{e}_{}^{2}{Q}_{}^{2}{r}_{}^{2}}{\Delta
}\right)\Phi = 0.
\end{eqnarray}                                         
   By decomposing \(\Phi\)  into \(\Phi ={e}_{}^{-i\omega
t}u(r)S(\theta ){e}_{}^{m\varphi} \)  ( here m is the angular momentum of the
particle parallel to the black hole rotating axis and \(\omega\) is the energy
of the particle ) and separating the variables, equation (26) is then reduced 
   to the equation governing the angular function  
\begin{equation}
\frac{1}{\sin \theta } \frac{d}{d\theta } \sin \theta  \frac{d}{d\theta }
S(\theta )+\left[{\lambda-{\left(\omega a\sin \theta - \frac{m}{\sin \theta }
\right)}_{}^{2}-\mu {a}_{}^{2}{\cos }_{}^{2}\theta} \right] S(\theta ) = 0 
\end{equation}       
  and the one governing the radial function 
\begin{equation}
\Delta  \frac{{d}_{}^{2}}{d{r}_{}^{2}}u(r)+2(r-M) \frac{d}{dr}u(r)=(\lambda
+{\mu }_{0}^{2}{r}_{}^{2} - \frac{{K}_{}^{2}}{\Delta } )u(r)=0, 
\end{equation}  
   where
\begin{equation}
K=({r}_{}^{2}+{a}_{}^{2})\omega -am-eQr
\end{equation}                   
   and \(\lambda \)  is a constant.                              
   If we use the tortoise coordinates transformation:
\begin{equation}
{r}_{\star }^{}= \pm \left[r+ \frac{1}{2{\kappa }_{+}^{}}\ln 
\frac{|r-{r}_{+}^{}| }{{r}_{+}^{}}- \frac{1}{2{\kappa }_{-}^{}}\ln 
\frac{|r-{r}_{-}^{}| }{{r}_{-}^{}}\right],
\end{equation}        
   where '+' is for the region 
\(r>{r}_{+}^{}\) and ' - ' is for the region\(r<{r}_{-}^{}\)  , function (28)
will be reduced to     
\begin{equation}
{({r}_{}^{2}+{a}_{}^{2})}_{}^{2} \frac{{d}_{}^{2}}{d{r}_{\star}^{2
}}u({r}_{\star}^{})+2r\Delta\frac{d}{d{r}_{\star}^{}}u({r}_{\star}^{
}) = \left[\Delta (\lambda +{\mu}_{0}^{2}{r}_{}^{2})-{K}_{}^{2}\right]u({r}_{\star}^{})=0.  
\end{equation} 
    When\(r\rightarrow {r}_{\pm}^{} \)  , we have\(\Delta \rightarrow 0\)  , so function (31)
is reduced to 
\begin{equation}
 \frac{{d}_{}^{2}}{d{r}_{\star}^{2}} u({r}_{\star}^{}) +
\frac{{K}_{}^{2}}{{({r}_{\pm}^{2}+{a}_{}^{2})}_{}^{2}}u({r}_{\star}^{})=0,
 \end{equation}    
  which equals to
\begin{equation}
\frac{{d}_{}^{2}}{d{r}_{\star}^{2}} u({r}_{\star}^{}) +{({\omega
}_{}^{2}-{\omega }_{\pm}^{})}_{}^{2}u({r}_{\star}^{})=0, 
\end{equation} 
   where \({\omega }_{\pm}^{
}=m{\Omega }_{\pm}^{}+e{V}_{\pm}^{}, {\Omega }_{\pm}^{}=
\frac{a}{{r}_{\pm}^{2}+{a}_{}^{2}}, {V}_{\pm}^{}=
\frac{Q{r}_{\pm}^{}}{{r}_{\pm}^{2}+{a}_{}^{2}}.\) 
 
 \({\Omega }_{\pm}^{} \)  are the angular velocities at the outer and inner
horizon and \({V}_{\pm}^{} \)  are the static electric potentials on the two
polar points (\(\theta =0 and \pi \)) at the outer and inner horizon respectively.     
When \(r\rightarrow {r}_{-}^{}\) , the
solution to the radial function (33) is      
\begin{equation}
u={e}_{}^{\pm i(\omega-{\omega }_{-}^{}){r}_{\star}^{}},
\end{equation}            
  so we have solved the outgoing wave        
\begin{equation}
{u}_{}^{out}={e}_{}^{-i\omega t+i(\omega-{\omega }_{-}^{}){r}_{\star}^{}
}={e}_{}^{-i\omega \nu }      (r<{r}_{-}^{})
\end{equation}
  and the ingoing wave 
\begin{equation}
{u}_{}^{in}={e}_{}^{-i\omega t-i(\omega-{\omega }_{-}^{}){r}_{\star}^{}
}={e}_{}^{-i\omega \nu } \cdot {e}_{}^{-2i(\omega -{\omega
}_{-}^{}){r}_{\star}^{}}        (r<{r}_{-}^{}) ,
 \end{equation}        
where we use the retarded Eddington coordinate:                  
\begin{equation}
\nu = t-\frac{\omega-\omega_{}^{-}}{\omega}{r}_{\star}^{}.
 \end{equation}              
   When \(r\rightarrow {r}_{-}^{}\) , we
see from the tortoise coordinates transformation (30) that   
\begin{equation}
{r}_{\star}^{}\rightarrow  \frac{1}{2{\kappa }_{-}^{}}\ln ({r}_{-}^{}-r), 
\end{equation}                 
  so solution (36) is reduced to                 
\begin{equation}
{u}_{}^{in}={e}_{}^{-i\omega \nu }\cdot {({r}_{-}^{}-r)}_{}^{-i(\omega
-{\omega }_{-}^{})/{\kappa }_{-}^{}}.
 \end{equation}  
     We then extend the solution (39) to the region \(  r\ge {r}_{-}^{}
\)  by analytic continuation:   
\begin{eqnarray}
{u}_{}^{in}=L({r}_{-}^{}-r){e}_{}^{-i\omega \nu}\cdot
{({r}_{-}^{}-r)}_{}^{-i(\omega -{\omega }_{-}^{})/{\kappa
}_{-}^{}}  \nonumber \\
+L(r-{r}_{-}^{}){e}_{}^{-i\omega\nu +\pi (\omega-{\omega
}_{-}^{})/{\kappa }_{-}^{}\cdot {(r-{r}_{-}^{})}_{}^{-i(\omega-{\omega}_{-}^{})/{\kappa }_{-}^{}}}, 
 \end{eqnarray}                           
    where                      
\begin{equation}
L(r) = \{ \begin{array}{lccr} 1 & r\ge 0 & & \\ 0 & r<0 & &
\end{array}  ,
\end{equation}                        
   so the overall ingoing wave could be written as                         
\begin{eqnarray}
{u}_{\omega }^{}={N}_{\omega }^{}\{ L({r}_{-}^{}-r){e}_{}^{-i\omega \nu }\cdot
{({r}_{-}^{}-r)}_{}^{-i(\omega -{\omega }_{-}^{})/{\kappa }_{-}^{}} \nonumber
\\ 
+L(r-{r}_{-}^{}){e}_{}^{-i\omega \nu+\pi (\omega -{\omega }_{-}^{})/{\kappa
}_{-}^{}}\cdot {(r-{r}_{-}^{})}_{}^{-i(\omega -{\omega }_{-}^{})/{\kappa
}_{-}^{}}\} ,
\end{eqnarray}                                               
   where  \({N}_{\omega}^{}\)is a normalization factor,
and \({N}_{\omega}^{2}\)  stands for the spectrum of radiation.  
    Because \(({u}_{\omega}^{},{u}_{\omega}^{})={N}_{\omega}^{2
}(1\pm{e}_{}^{2\pi(\omega -{\omega }_{-}^{})/{\kappa }_{-}^{} } )=\pm 1,\)
where '+ ' is for fermions and ' - ' is for bosons , so we obtain              
\begin{equation}
{N}_{\omega}^{2}= \frac{1}{{e}_{}^{(\omega -{\omega
}_{-}^{})/{T}_{-}^{}}\pm 1}, 
 \end{equation}                            
\begin{equation}
{T}_{-}^{}= \frac{{\kappa }_{-}^{}}{2\pi }.
\end{equation}                                
    Thus we proved that the inner horizon of the Kerr-Newman black
hole has the temperature \({T}_{-}^{}= \frac{{\kappa }_{-}^{}}{2\pi }
\) , and there are particles generated by vacuum polarization erupted from   the
inside (the region \(r<{r}_{-}^{}\) ) to the inner horizon. This makes a new
kind of radiation ( or absorption[24]). The particle that arrives at the inner horizon will travel
to the outer horizon and reduce its temperature to \({T}_{+}^{}=
\frac{{\kappa }_{+}^{}}{2\pi } \). The particle then goes
ahead to erupt from the outer horizon, which makes the usual Hawking
radiation.        
   The Bekenstein-Smarr equation of the black hole could be written simply as formula (4).
Thus we get \(S= \frac{A}{4}\), where A is the area of outer horizon. Also, the 
Bekenstein-Smarr equation could be written in anothor form:
\begin{equation}
dM= \frac{{\kappa }_{-}^{}}{8\pi}d{A}_{-}^{}+{\Omega}_{-}^{} dJ+{V}_{-}^{}dQ,
 \end{equation}    
  where \({A}_{-}^{}\) 
 is the area of inner horizon :                       
\begin{equation}
{A}_{-}^{}=-\- 4\pi ({r}_{-}^{2}+{a}_{}^{2}).
\end{equation}     
  We can assume that the contribution to
the entropy from the outer and inner horizon are \({S}_{+}^{}\) 
and \({S}_{-}^{}\)  respectively:  
\begin{equation}
{S}_{\pm}^{}= \frac{{A}_{\pm}^{}}{4},
\end{equation}                                   
   so the total entropy is              
\begin{equation}
S={S}_{+}^{}+{S}_{-}^{}=4\pi M \sqrt{{M}_{}^{2}-{a}_{}^{2}-{Q}_{}^{2} }, 
\end{equation}
which is the same with formula (14), and obviously,  
vanishes as temperature vanishes. we therefore think the entropy of the extremal
Kerr-Newman black hole is zero and the third law of thermodynamics is still
applicable in this case.


\begin{thebibliography}{999}
\bibitem{1} J.M.Bardeen,B.Carter,S.W.Hawking, {\em Commun.Math.Phys.} {\bf
31} (1973) 161.
\bibitem{2} S.W.Hawking, {\em Commun.Math.Phys.\/} {\bf 25} (1972) 152.

\bibitem{3} J.Bekenstein, {\em Ph.D. Thesis \/} ( {\em Princeton Uniersity \/}
, 1972 ).  

\bibitem{4} L.Smarr, {\em Phys.Rev.Lett.\/} {\bf 30} (1973) 71.
  

\bibitem{5} S.W.Hawking, {\em Commun.Math.Phys.\/} {\bf 43} (1975) 199.

\bibitem{6} W.G.Unruh , R.M.Wald, {\em Phy.Rev.D\/}  {\bf 25} (1982) 942.
 

\bibitem{7} V.P.Frolov , D.N.Page , {\em Phys.Rev.Lett.\/}  {\bf 71} (1993)  
3902.  

\bibitem{8} R.M.Wald ,{\em Quantum Field Theory in Curved Space-Time and Black
Hole Thermodynamics \/}({\em University of Chicago Press,Chicago\/}s,1994).
 
\bibitem{9} V.P.Frolov, I.D.Novikov, {\em Black Hole Physics \/} ( {\em
Kluwer Academic Publisher , Netherlands ,\/} 1998 ).
  
\bibitem{10} R.M.Wald, {\em Phy.Rev.D \/} {\bf 56} (1997) 6467 . 

\bibitem{11} H. Lin, {\em Acta Physica Sinica\/} {\bf 49-8} (2000) 1413 . 

\bibitem{12} S.W.Hawking , G.Horowitz , S.Ross ,{\em
Phy.Rev.D \/} {\bf 51} (1995) 4302 .

\bibitem{13} G.W.Gibbons , R.E.Kallosh, {\em Phy.Rev.D\/} {\bf
51} (1995) 2839 .

\bibitem{14} C.Teitelbom ,{\em Phy.Rev.D \/} {\bf 51} (1995) 4315 . 

\bibitem{15} A.Ghosh ,P.Mitra,{\em Phys.Rev.Lett.\/} {\bf 77} (1996) 4848. 

\bibitem{16} B. Wang, R. S. Su, {\em Phys. Lett. } {\bf B 432} (1998) 69, gr-qc/9807050.  

\bibitem{17} A.Ghosh,P.Mitra,{\em Phys.Rev.Lett.\/} {\bf 78} (1996) 1858.

\bibitem{18} D.J.Loranz,et al.,{\em Phy.Rev.D \/} {\bf 52} (1995) 4554. 

\bibitem{19} O.B.Zaslavskii, {\em Phys.Rev.Lett.\/} {\bf 76} (1996)2211. 

\bibitem{20} O.B.Zaslavskii, {\em Phy.Rev.D \/} {\bf 56} (1997)2188. 

\bibitem{21} O.B.Zaslavskii, {\em Phy.Rev.D \/} {\bf 56} (1997)6695.

\bibitem{22} D.J.Konkowski ,T.M.Helliwell, {\em Phy.Rev.D \/}
{\bf 54} (1996) 7898.   

\bibitem{23} J.M.Maldacena,A.Strominger, {\em Phys.Rev.Lett.\/}
{\bf 77} (1996) 430,  hep-th/9603195. 

\bibitem{24} C.G. Callan, S.S. Gubser, I.R. Klebanov, A.A. Tseytlin, {\em Nucl.Phys.\/}
{\bf B489} (1997) 65, hep-th/9610172.

\end{thebibliography}
\end{document}